\documentclass[prd,aps,preprint,tightenlines,nofootinbib,superscriptaddress]{revtex4}
\usepackage{mathrsfs}
\usepackage{amsfonts}
\usepackage{amsmath}
\usepackage{array}
\usepackage{verbatim}
\usepackage{epsfig}
\usepackage{graphicx}

\begin{document}
\title{Long Range Correlation in Higgs Boson Plus Two Jets Production at the LHC}

\author{Peng Sun}
\affiliation{Nuclear Science Division, Lawrence Berkeley National
Laboratory, Berkeley, CA 94720, USA}
\author{C.-P. Yuan}
\affiliation{Department of Physics and Astronomy, Michigan State University,
East Lansing, MI 48824, USA}
\author{Feng Yuan}
\affiliation{Nuclear Science Division, Lawrence Berkeley National
Laboratory, Berkeley, CA 94720, USA}

\begin{abstract}
We study Higgs boson plus two high energy jets production at the LHC in the kinematics
where the two jets are well separated in rapidity. The partonic processes are dominated
by the $t$-channel weak boson fusion (WBF) and gluon fusion (GF) contributions. 
We derive the associated QCD resummation formalism for the correlation analysis
where the total transverse momentum $q_\perp$ of the Higgs boson 
and two jets is small. Because of different color structures, the resummation results 
lead to distinguished behaviors: the WBF contribution peaks at relative low
$q_\perp$ while all GF channel contributions are strongly de-correlated and spread
to a much wider $q_\perp$ range.
By applying a kinematic cut on $q_\perp$, one can effectively increase 
the WBF signal to the GF background by a significant factor.
This greatly strengthens the ability to investigate the WBF channel in Higgs 
boson production and study the couplings of Higgs to electroweak bosons. 
\end{abstract}

\maketitle

{\it Introduction.} 
One of the most important physics tasks after the discovery of the Standard Model (SM) Higgs 
boson at the CERN LHC~\cite{Aad:2012tfa,Chatrchyan:2012ufa} is to investigate the
coupling between the Higgs boson and the SM particles, in particular, the electroweak
bosons. An important channel to study this coupling is through the Higgs boson plus 
two jets production with large rapidity separation between the jets, 
where the weak-boson fusion (WBF) contribution dominates over the 
gluon fusion (GF) contribution~\cite{Aad:2014lwa,Aad:2014tca,Aad:2014eha,Campbell:2006xx,
Campbell:2012am,Figy:2003nv,Ellis:2005qe,Gangal:2013nxa,Dittmaier:2012vm}. It has been further argued that
because of colorless exchange in the WBF contribution as compared to color exchange
in the GF contribution, they can be discriminated by the correlation study
between the Higgs boson and the two jets. By imposing additional kinematic requirements
to reflect the above feature will help to enhance the WBF signal to the GF background ratio. 
In this paper, we will demonstrate that the total transverse momentum of the Higgs boson 
and the two jets can be used as an important probe to distinguish the WBF and
GF mechanisms. 

Higgs boson plus two jets are produced in $pp$ collisions at the LHC through
\begin{equation}
A(P )+B(\bar P)\to H+Jet_1+Jet_2+X \ ,
\end{equation}
where the incoming nucleons carry momenta $P$ and $\bar P$, and the final
state Higgs boson and the two jets with momenta $P_h$, $k_1$ and $k_2$,
respectively. We are interested in the kinematics that one jet is 
produced in the forward direction and another jet in the backward direction,
while the Higgs boson in the central region.
Because of the large rapidity difference between the two final state jets, 
i.e., $\Delta y_{12}=|y_{j1}-y_{j2}|\gg 0$, 
the above process is dominated by the $t$-channel weak
boson or gluon exchange diagrams for WBF or GF contributions, 
respectively.
In the correlation kinematics, the total transverse
momentum $\vec{q}_\perp=\vec{P}_{h\perp}+\vec{k}_{1\perp}+\vec{k}_{2\perp}$
is small. The leading order diagrams $a(p_1)+b(p_2)\to c(k_1)+H(P_h)+d(k_2)$ 
contribute to a Delta function of $\delta^{(2)}(q_\perp)$.
Due to different color structures, the WBF and GF channels will 
lead to different $q_\perp$ distributions from higher order corrections. 
This will provide additional handle to differentiate 
the WBF and GF contributions in Higgs boson plus two jets production.
However, the fixed order perturbative corrections 
lead to a singular distribution at low $q_\perp$. Therefore, in order
to consolidate the powerful reach of the correlation study,
we need to include an all order resummation. The goal
of this paper is to derive the associated QCD resummation formalism
and to demonstrate the powerful probe discussed above.

The QCD resummation in the low $q_\perp$ region is referred as the transverse 
momentum dependent (TMD) resummation or the Collins-Soper-Sterman (CSS) 
resummation~\cite{Collins:1984kg}. In our study, 
we follow the CSS procedure and apply recent developments on the 
TMD resummation for jet production in the final 
state~\cite{Banfi:2008qs,Mueller:2013wwa,Sun:2014gfa,Sun:2014lna}.
An important feature of our derivation is the special
kinematics mentioned above with $\Delta y_{12}\gg 0$ where the final
state radiation associated with the jets can be resummed through a simple
soft factor. The final resummation formula can be summarized
into the following form, up to next-to-leading logarithmic (NLL) order,
\begin{eqnarray}
\frac{d^4\sigma}
{dy_h dy_{j1} dy_{j2}d k_{1\perp}^2d k_{2\perp}^2
d^2q_{\perp}}|_{resum.}=\sum_{ab}\sigma_0\int\frac{d^2\vec{b}_\perp}{(2\pi)^2}
e^{-i\vec{q}_\perp\cdot
\vec{b}_\perp}W_{ab\to cHd}(x_1,x_2,b_\perp) \ ,
\end{eqnarray}
where 
$\sigma_0$ represents the normalization of the differential cross section
from the leading order diagrams.
An all order resummation of $W(b_\perp)$ is written as 
\begin{eqnarray}
W\left(x_1,x_2,b_\perp\right)&=&{\cal H}(\hat \mu)x_1f_a(x_1,\mu_b)
x_2f_b(x_2,\mu_b) e^{-S_a(\hat\mu,b_\perp)-S_b(\hat\mu,b_\perp)} \ ,\label{resum}
\end{eqnarray}
where $H$ represents the hard coefficients depending on factorization
scale $\hat\mu$, $\mu_b=b_0/b_\perp$ with $b_0=2e^{-\gamma_E}$, 
$f_{a,b}(x,\mu_b)$ are parton distributions for the incoming 
partons $a$ and $b$, and $x_{1,2}$ are momentum fractions of 
the incoming hadrons carried by the partons. 
($\gamma_E$ is the Euler's constant.) 
The two Sudakov form factors collect contributions from soft gluon radiations 
involved in both the initial and final states of a given partonic process, 
specified by the two incoming partons ($a$ and $b$) of the colliding nucleons.  
 For the parton ``$a$", we have
\begin{eqnarray}
S_{a}(\hat\mu,b_\perp)=\int^{\hat\mu^2}_{\mu_b^2}\frac{d\mu^2}{\mu^2}
\left[\ln\left(\frac{s}{\mu^2}\right)A_a+B_a+D_a\ln\frac{1}{R^2}+\gamma_a^{\prime s}\right]\ , \label{su}
\end{eqnarray}
where $s=(p_1+p_2)^2$ is the total partonic center of mass energy
squared, $R$ the jet size, and $A$, $B$ and $D$ are perturbative coefficients, e.g.,
$A=\sum A^{(i)}(\alpha_s/2\pi)^i$.
$A$ and $B$ are the same as those for the inclusive $Z$ boson
and Higgs boson production, via quark fusion and gluon fusion processes, respectively, 
with $A_q^{(1)}=C_F$, $A_g^{(1)}=C_A$,
$B_q^{(1)}=-\frac{3}{2}C_F$, $B_g^{(1)}=-2\beta_0C_A$ where $\beta_0=11/12-N_f/18$. 
Because of the soft gluon radiation contributions associated with the final state jets, we have
additional coefficients $D_q^{(1)}=C_F$ and $D_g^{(1)}=C_A$ for quark and gluon
jet, respectively. (In QCD, $C_F=\frac{4}{3}$, $C_A=3$, and $N_f$ is the number 
of light quark flavors at a given energy scale.) The last term is the most important
term in our calculation, because it further discriminates the WBF and 
GF channels in the Higgs boson plus two jets production processes. 
We find for each incoming parton ``$a$'', depending on either WBF or GF production mechanism,  
\begin{equation}
\gamma_{qWBF}^{\prime s}=-C_F\ln\frac{u_1}{t_1},~~
\gamma_{qGF}^{\prime s}=(C_A-C_F)\ln\frac{u_1}{t_1},~~
\gamma_{gGF}^{\prime s}=0 \ ,\label{gammas}
\end{equation}
where $t_1=-2k_1\cdot p_1$ and $u_1=-2k_1\cdot p_2$. Similar expressions hold
for parton ``$b$" but with $t_2=-2k_2\cdot p_2$ and $u_2=-2k_2\cdot p_1$.
In the kinematics we are interested in, i.e., $\Delta y_{12}\gg 0$ and $y_1y_2<0$,
we have the following relations $|u_1|\gg |t_1|$ and $|u_2|\gg |t_2|$.
From the above results, we can clearly see that the leading double logarithms are
universal among different channels, depending on the color charge of the 
incoming partons. However, the sub-leading logarithms differ among the 
WBF and GF channels. This additional term, 
proportional to $C_A\ln(u_1/t_1)$, will play a significant role 
to differentiate the WBF from GF processes in the proposed kinematical region, 
with large rapidity gap of the two forward jets, due 
to the fact that $|u_1|\gg |t_1|$.
In the following, we will briefly present the major steps to derive
the above resummation formula, and detailed derivations will be
presented in a separate publication. 

{\it Asymptotic behavior at low $q_\perp$.}
As shown in Fig.~1, the leading order contributions from both
WBF and GF channels lead to a Delta 
function in $q_\perp$. One gluon radiation will result into a singular 
behavior at low $q_\perp$. 
In the WBF channel, because of colorless exchange in the $t$-channel, 
there is no interference between the gluon radiation from the upper quark line and the lower quark line.
The only nontrivial part is how to deal with the jet contribution, where 
we need to exclude the soft gluon radiation contributing to the final state jet functions. 
This exclusion will naturally introduce the 
jet cone size dependence in the soft gluon radiation. 
Following the recent developments in Refs.~\cite{Mueller:2013wwa,Sun:2014gfa,Sun:2014lna}, we find
that at low $q_\perp$ the differential cross section can be written as
\begin{eqnarray}
&&\frac{\alpha_sC_F}{2\pi^2}\frac{1}{q_\perp^2}\int\frac{dx_1'dx_2'}{x_1'x_2'} x_1'f_q(x_1')x_2'f_{q'}(x_2')
\left[\left\{\delta(\xi_2-1)\xi_1{\cal P}_{qq}(\xi_1)+(\xi_1\leftrightarrow \xi_2)\right\}\nonumber\right.\\
&&\left.+\delta(\xi_1-1)\delta(\xi_2-1)\left(2\ln\frac{s}{q_\perp^2}+\ln\frac{t_1t_2}{u_1u_2}
-3+2\ln\frac{1}{R^2}\right)\right] \ ,\label{softv}
\end{eqnarray} 
for the WBF contribution, where ${\cal P}_{qq}$ is the quark splitting kernel. 
To derive the jet size dependence, we have applied the narrow jet approximation and the 
anti-$k_t$ algorithm~\cite{Jager:2004jh}.
However, for the GF contribution in quark-quark scattering channel, the color structure is different from the
WBF contribution. In addition to the terms in Eq.~(\ref{softv}), 
the interference between the quark lines
gives the following additional term,
\begin{eqnarray}
\frac{\alpha_s}{2\pi^2}\frac{C_A}{q_\perp^2}\ln\frac{u_1u_2}{t_1t_2}\ ,\label{softg}
\end{eqnarray} 
whose contribution becomes large when the rapidity difference between the two 
final state jets is large, namely, when $|u_1| \gg |t_1|$ or $|u_2| \gg |t_2|$.   
Similar results can be obtained for the gluon-gluon and quark-gluon 
scattering channels.

\begin{figure}[tbp]
\centering
\includegraphics[width=10cm]{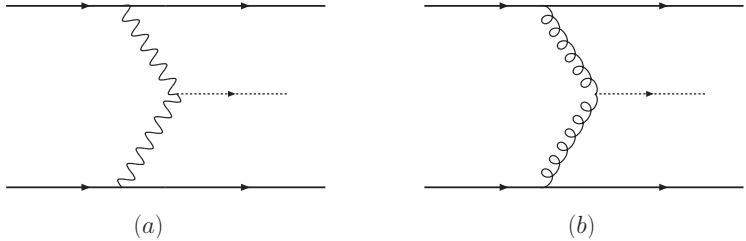}
\caption{Two production mechanism for Higgs plus two jets in the quark-quark
scattering channel: (a) vector boson fusion and (b) gluon fusion.}
\label{lo}
\end{figure}

When Fourier transformed into the $b_\perp$ space to
calculate the one-loop corrections to $W(b_\perp)$, c.f. Eq. (2), 
the above results will contain soft
divergences ($1/\epsilon^2$ in dimension regulation with $D=4-2\epsilon$ dimension). 
These soft divergences will be canceled out by the virtual diagrams. 
This provides an important cross check of the above results in the low $q_\perp$ region. 
The total result of real and virtual contributions will be also used
to demonstrate the TMD factorization in the following
section.

{\it TMD Factorization.}
We follow the Collins 2011 formalism~\cite{Collins:2011zzd}, where the TMD
parton distributions are defined with soft factor subtraction~\footnote{Different
schemes can be applied to the TMDs, which will lead to the same final 
resummation formula in the CSS framework~\cite{Catani:2000vq,Ji:2004wu,Prokudin:2015ysa}.}. For example, the
TMD quark distribution is defined as~\cite{Collins:2011zzd},
\begin{equation}
f_{q}^{sub.}(x,b_\perp,\hat\mu,\zeta_c)=f_q^{unsub.}(x,b_\perp)\sqrt{\frac{S^{\bar
n,v}(b_\perp)}{S^{n,\bar n}(b_\perp)S^{n,v}(b_\perp)}} \ , \label{jcc}
\end{equation}
where $b_\perp$ is the Fourier conjugate variable respect to
the transverse momentum $k_\perp$, $\hat \mu$ the factorization
scale and $\zeta_c^2=x^2(2v\cdot P)^2/v^2=2(xP^+)^2e^{-2y_n}$ with
$y_n$ the rapidity cut-off in Collins-11 scheme. The second factor represents the 
soft factor subtraction with $n$ and $\bar n$ as the light-front vectors $n=(1^-,0^+,0_\perp)$,
$\bar n=(0^-,1^+,0_\perp)$, whereas $v$ is an off-light-front vector, and
$v=(v^-,v^+,0_\perp)$ with $v^-\gg v^+$. 
The un-subtracted TMD quark distribution reads as 
 \begin{eqnarray}
f_q^{unsub.}(x,k_\perp)&=&\frac{1}{2}\int
        \frac{d\xi^-d^2\xi_\perp}{(2\pi)^3}e^{-ix\xi^-P^++i\vec{\xi}_\perp\cdot
        \vec{k}_\perp}  \left\langle
PS\left|\overline\psi(\xi){\cal L}_{n}^\dagger(\xi)\gamma^+{\cal L}_{n}(0)
        \psi(0)\right|PS\right\rangle\ ,\label{tmdun}
\end{eqnarray}
with the gauge link defined as $ {\cal L}_{n}(\xi) \equiv \exp\left(-ig\int^{-\infty}_0 d\lambda
\, v\cdot A(\lambda n +\xi)\right)$. The light-cone singularity in the un-subtracted
TMDs is cancelled out by the soft factor, as in Eq.~(\ref{jcc}), with $S^{v_1,v_2}$
defined as
\begin{equation}
S^{v_1,v_2}(b_\perp)={\langle 0|{\cal L}_{v_2}^\dagger(b_\perp) {\cal
L}_{v_1}^\dagger(b_\perp){\cal L}_{v_1}(0){\cal
L}_{v_2}(0)  |0\rangle   }\, . \label{softg}
\end{equation}
Similarly, we can define the TMD gluon distribution function.

The above TMD distribution functions (TMDs) are defined for the hard processes with color neutral particle
production in the final state. To apply these TMDs in our process, we have to also
include the soft gluon radiation from the final state jets. 
In previous studies of dijet production, the additional soft factor was expressed
in the matrix form~\cite{Sun:2014gfa}. However, in the current case, the two jets are 
produced with large rapidity separation, and the leading contribution comes from 
the $t$-channel diagrams. Because of that, the soft factor can
be simplified as
\begin{eqnarray}
{\bf S}_{(\bar 1,\bar 8)}(b_\perp,R;\hat \mu)&=&\int_0^\pi
\frac{d\phi_0}{\pi}\; \frac{C^{bb'}_{Iii'} C^{aa'}_{Ill'}}{S^{n,\bar n}(b_\perp)}
\langle 0|{\cal L}_{ncb'}^\dagger(b_\perp) {\cal
L}_{ nbc} (b_\perp){\cal L}_{\bar
nca'}^\dagger(0) {\cal L}_{\bar nac}(0) \nonumber\\
&&\times {\cal L}_{n_1 ji}^\dagger(b_\perp) {\cal
L}_{n_2 i'k}(b_\perp) {\cal L}_{n_2kl}^\dagger (0) {\cal
L}_{n_1l'j} (0)  |0\rangle \ ,\label{soft}
\end{eqnarray}
in the quark-quark scattering channel from either the color-singlet $(\bar 1)$ for the WBF contribution, 
or the color-octet $(\bar 8)$ for the GF contribution, 
respectively. To project out the color-singlet contribution we take
$C^{bb'}_{\bar 1ii'}=\delta_{bb'}\delta_{ii'}$, whereas $C^{bb'}_{\bar 8ii'}=T^e_{bb'}T^e_{ii'}$
for the color-octet case. Again, we have applied the subtraction method to define the
soft factor, where the light-cone singularity from the gauge links associated with
the incoming partons are cancelled out. In addition, we average out the azimuthal angle $\phi_0$ of the
leading jet and retain the relative azimuthal angle $\phi$ for $q_\perp$, where
$n_1$ and $n_2$ represent final state two quark jets' momentum
directions. In deriving the soft factor ${\bf S}(b_\perp)$, we need to exclude soft gluon radiation contributing to the 
final state jet function, which falls inside the jet and leads to the jet size ($R$) dependence in the
soft factor.

In the end, the TMD factorization for $W(b_\perp)$ can be written as
\begin{eqnarray}
W(b_\perp)&=&f_q(x_1,b_\perp,\zeta_c;\hat\mu)f_q(x_2,b_\perp,\zeta_c';\hat\mu)
{\bf S}_{(\bar 1,\bar 8)}(b_\perp;\hat\mu){\cal H}_{TMD}(\hat \mu;s,t_1,u_1,t_2,u_2) \ , \label{tmd}
\end{eqnarray}
for quark-quark scattering channel from WBF $(\bar 1)$ and GF $(\bar 8)$ contributions,
respectively.
The TMD quark distributions are the same for both WBF and GF production mechanisms, 
whereas the soft and hard factors are different. The explicit calculations at one-loop order verify 
the above factorization formula in terms of the TMDs. We are left with the finite
contributions for the hard factor ${\cal H}_{TMD}$. For WBF channel, we find that at the 
next-to-leading order (NLO), 
\begin{eqnarray}
{\cal H}_{TMD}^{WBF}&=&1+\frac{\alpha_s}{2\pi}C_F\left\{
\left[\frac{1}{2}\ln^2\left(\frac{k_{1\perp}^2}{\hat\mu^2}\right)-\ln\frac{k_{1\perp}^2}{\hat\mu^2}
\left(2\ln\frac{-t_1}{k_{1\perp}^2}+\ln\frac{1}{R^2}-\frac{3}{2}\right)
-\ln^2\left(\frac{-t_1}{k_{1\perp}^2}\right)\right.\right.\nonumber\\
&&\left.\left.+3\ln\frac{-t_1}{k_{1\perp}^2}+\frac{3}{2}\ln\frac{1}{R^2}-\frac{3}{2}-\frac{5\pi^2}{6}\right]
+\left(t_1\leftrightarrow t_2,k_{1\perp}\leftrightarrow k_{2\perp}\right)\right\} \ ,
\end{eqnarray}
where we have taken $\zeta_c^2=\zeta_c^{\prime 2}=s$ to simplify the 
final results. For the GF contribution, the hard factor is too lengthy to 
be listed here.  We emphasize, however, that the similar logarithmic terms appear.
We have also verified the TMD factorization for the gluon-gluon and
quark-gluon channels, with the TMD gluon distributions from the incoming
nucleons and the associated soft factors. 
 
{\it Resummation and Phenomenological Applications.} The large logarithms are 
resummed by solving the relevant evolution equations for the individual factors
in the TMD factorization formula in Eq.~(\ref{tmd}). For example, the TMD parton 
distributions obey two evolution equation:
one is associated with the rapidity cut-off parameter $\zeta_c$ and one associated
with the factorization scale $\hat\mu$~\cite{Collins:2011zzd}. 
Additional resummation of large logarithms are carried out by solving the renormalization
group equation for the soft factor, which is controlled by the associated 
anomalous dimension,
\begin{equation}
\frac{\partial}{\partial\ln\hat\mu}{\bf S}_{\bar 1,\bar 8}(b_\perp;\hat\mu)=
-2\gamma_{\bar 1,\bar 8}^{s}{\bf S}_{\bar 1,\bar 8}(b_\perp;\hat\mu) \ .
\end{equation}
From the one-loop calculations, we find the following results for the anomalous dimension
for the soft factors in all partonic channels $ab\to cHd$,
\begin{equation}
\gamma_{\bar 1,\bar 8}^{s}=\frac{\alpha_s}{2\pi}\left[(D_a+D_b)\ln\frac{1}{R^2}
+\gamma_a^{\prime s}+\gamma_b^{\prime s} \right]\ ,
\end{equation}
where $\gamma_{a,b}^{\prime s}$ have been given in Eq.~(\ref{gammas})
and $D_{a,b}$ defined as before. 
The final resummation formulas of Eqs.~(2)-(5) are obtained by solving the
above mentioned evolution equations. Choosing the factorization 
scale $\hat\mu\approx k_{1\perp}\sim k_{2\perp}$ will reduce the large logarithms
in the hard factors.

We would like to emphasize that the resummation formulas of Eqs.~(2)-(5) for
the WBF contribution are also valid in all rapidity regions for the final state
Higgs boson and the two jets. This is because, there are only $t$-channel 
color-less exchange diagrams contributing to the WBF process, and the 
relevant derivations above work for all kinematics. This is very much similar
to the structure function approach for the inclusive cross section for Higgs boson production
through WBF channels studied in Ref.~\cite{Han:1992hr}.

\begin{figure}[tbp]
\centering
\includegraphics[width=7cm]{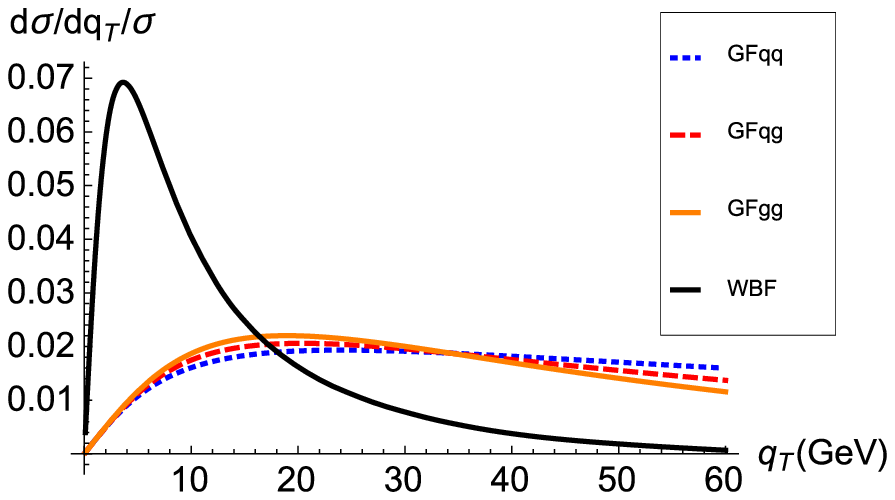}
\includegraphics[width=7cm]{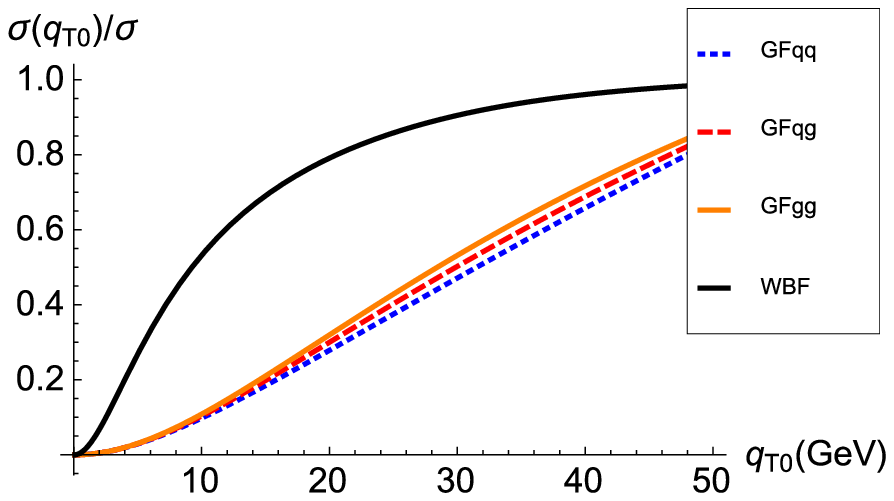}
\caption{Noramlized distributions of the vector boson fusion and gluon-fusion
contributions to the Higgs boson plus two jets production in the typical kinematics at the
LHC with $\sqrt{S}=13TeV$, where the jet transverse momenta 
$k_{1\perp}=k_{2\perp}=30GeV$, $y_{j1}=-y_{j2}=2$ and $y_h=0$:
as functions of  the total transverse momentum $q_\perp$ (left); 
the total rate as function of the upper limit of $q_\perp$ (right).}
\label{qtdistribution}
\end{figure}

In the following, we will apply the resummation formulas of Eqs.~(2)-(5) to 
calculate the $q_\perp$ distribution via either WBF or GF mechanisms, and 
show how to use the $q_\perp$ distribution to enhance the WBF and suppress the 
GF contributions. This is important for further testing the Standard Model prediction 
of the couplings of Higgs boson to weak gauge bosons. 
For illustration purpose, we examine the case that 
the Higgs boson is produced in the central rapidity region 
with $y_h=0$, while the two jets are in forward or backward 
rapidity regions with $y_{j1}=2$ and $y_{j2}=-2$. 
For the numeric calculations, we use the CT10 NLO parton distribution functions~\cite{Gao:2013xoa},
and take the Sudakov resummation coefficients listed in the Introduction.
The non-perturbative form factors are adopted from a recent study in Ref.~\cite{Su:2014wpa}.
We have also checked other existing parameterizations (e.g., those in Ref.~\cite{Sun:2012vc})
and found negligible difference in the numerical results.
In the left panel of Fig.~\ref{qtdistribution}, we plot the normalized distributions as functions of $q_\perp$ 
separately in the WBF and three different GF production channels: 
$\frac{1}{\sigma^{tot}}\frac{d\sigma}{dq_\perp}$ with $\sigma^{tot}$ obtained by integrating 
over $q_\perp$ from $0$ up to $60\,$ GeV in each channel. 
Clearly, we find that the WBF contribution peaks around 5-7 GeV, whereas the
GF contributions (via $qq$, $qg$ or $gg$ scattering processes)  produce much 
wider $q_\perp$ distributions. This is a direct consequence of the
difference in the Sudakov form factor coefficients associated with 
different scattering processes, cf. Eq. (5). 
When the two jets are produced with large rapidity separation, we have $|u_1u_2|\gg |t_1t_2|$,
and the difference between $\gamma_{qGF}^{\prime s}$ and
$\gamma_{qWBF}^{\prime s}$ leads to a significant broadening in the 
$q_\perp$ distribution for the GF contribution as compared to the 
WBF contribution. It is also interesting to notice that all the
GF channels  (via $qq$, $qg$ or $gg$ scattering processes) have similar distributions,
despite the fact that the additional term $\gamma^{'s}_{gGF}$, associated with the incoming 
gluon in the  ( $qg$ or $gg$)  GF processes vanishes, cf. Eq. (5). This is because 
$A_g$ is proportional to $C_A$ and the double log term dominates the Sudakov factor 
$S_g(\hat \mu,b_\perp)$ associated with the 
incoming gluon to result in similar $q_\perp$ distributions in all three GF processes, as shown in Fig.~2.
Physically, they all come from $t$-channel gluon
exchange, and gluon radiation in the large rapidity interval between the two
jets generate large Sudakov effects.

The dramatical difference in the $q_\perp$ distributions of the Higgs boson plus two jet 
system in the WBF and GF processes provides an important tool to distinguish these two production 
mechanisms. To further demonstrate this point, in the right panel of Fig.~\ref{qtdistribution}
we plot the ratio of the integrated cross section over $\sigma^{tot}$ as functions of the upper 
limit of the integration, denoted as $q_{\perp 0}$ there. 
Integrated up to 20 GeV, the WBF contribution is already at $80\%$ of the integrated cross section up to 60 GeV, 
while the GF contribution only reaches to $28\%$ of the integrated cross section of individual GF subprocess. 
Hence, we conclude that the predicted $q_\perp$ distributions can be used to further discriminate the production mechanisms of the Higgs boson plus two jets with large rapidity separation. In this kinematics, the WBF and GF production processes are characterized by the exchange of a colorless weak boson or a colored gluon, respectively. 
Requiring an upper limit in $q_\perp$ value will increase the fraction of the data sample induced by WBF, in contrast to GF, process.   
This will greatly benefit the detailed investigation
of the Higgs-electroweak boson coupling from this process.

{\it Summary and discussions.} 
In this paper, we have derived, for the first time, the QCD resummation
formula for the Higgs boson plus two jets production at the LHC, in the 
most interesting kinematics that the two jets are separated with large rapidity
difference. 
Explicit one-loop calculations for both WBF and GF contributions were
performed, and all order resummation formulas were obtained in the low
$q_\perp$ region of the total transverse momentum of
the Higgs boson and two jets. We have also demonstrated that 
an additional upper limit on $q_\perp$ will enhance the WBF signal as compared to the
GF background. 
The low $q_\perp$ region also corresponds to the back-to-back correlation region in the azimuthal
angular distribution between the Higgs boson and the two jets.
Further studies shall follow to combine our resummation formulas with the
existing codes for NLO calculations, such as MCFM~\cite{Campbell:2010ff} and
VBFNLO~\cite{Arnold:2008rz}, to have more detailed
phenomenological investigations. Theoretically, we should also pursue
the QCD resummation derivation for generic kinematics of Higgs boson
plus two jets production, where a matrix form will be required for the
GF contributions, similar to the dijet production process. 
For the WBF contribution, our resummation formulas,
Eqs. (2)-(5), can be applied to all the kinematic regions of Higgs 
boson plus two jets production in hadron collision.

From our derivations, we have shown that the GF contributions are dominated by
the $t$-channel gluon exchange diagrams, which generate a significant resummation
effects in terms of $\ln(u_1u_2)/(t_1t_2)$. These enhancements will increase
with rapidity difference $\Delta y_{12}$ between the two jets. In our calculations, we have
formulated these contributions as a TMD soft factor, and resummation
was carried out by following the CSS procedure. At very large rapidity separation,
we may have to consider the BFKL resummation, similar to that of the so-called
Mueller-Navelet dijet production~\cite{Mueller:1986ey,Mueller:2015ael}. 
How and when we should include BFKL dynamics is an important  
question that needs further investigations. 

Last, we would like to emphasize that the method developed in this paper can
be applied to the new physics search as well. Especially, for the new particle
production through the weak boson fusion processes, the 
resummation would be similar to the WBF contribution to the Higgs boson plus two 
jets production. Therefore, we can apply the same kinematic cut in $q_\perp$ to enhance
the new physics signal as compared to the QCD background.

This material is based upon work supported by the U.S. Department of Energy, 
Office of Science, Office of Nuclear Physics, under contract number 
DE-AC02-05CH11231, and by the U.S. National 
Science Foundation under Grant No. PHY-1417326.

\end{document}